# Formation of ultracold $^7$Li$^{85}$Rb molecules in the lowest triplet electronic state by photoassociation and their detection by ionization spectroscopy

Adeel Altaf[1,a)], Sourav Dutta[1,*], John Lorenz[1], Jesús Pérez-Ríos[1], Yong P. Chen[1,2] and D. S. Elliott[1,2,b)]

[1]*Department of Physics and Astronomy, Purdue University, West Lafayette, Indiana 47907, USA*

[2]*School of Electrical and Computer Engineering, Purdue University, West Lafayette, Indiana 47907, USA*

We report the formation of ultracold $^7$Li$^{85}$Rb molecules in the $a^3\Sigma^+$ electronic state by photoassociation (PA) and their detection via resonantly enhanced multiphoton ionization (REMPI). With our dual-species Li and Rb magneto-optical trap (MOT) apparatus, we detect PA resonances with binding energies up to ~62 cm$^{-1}$ below the $^7$Li 2s $^2S_{1/2}$ + $^{85}$Rb 5p $^2P_{1/2}$ asymptote. In addition, we use REMPI spectroscopy to probe the $a^3\Sigma^+$ state and excited electronic $3^3\Pi$ and $4^3\Sigma^+$ states, and identify $a^3\Sigma^+$ ($v'' = 7 - 13$), $3^3\Pi$ ($v_\Pi' = 0 - 10$) and $4^3\Sigma^+$ ($v_\Sigma' = 0 - 5$) vibrational levels. Our line assignments agree well with *ab initio* calculations. These preliminary spectroscopic studies on previously unobserved electronic states are crucial to discovering transition pathways for transferring ultracold LiRb molecules created via PA to deeply bound rovibrational levels of the electronic ground state.

## I. INTRODUCTION

Ultracold molecules are a source of great interest and a focus of active research due to their promise in areas of research such as novel quantum phase transitions [1], ultracold chemistry [2, 3, 4], tests of fundamental forces [5, 6, 7] and others. A subset of these ultracold molecules is heteronuclear diatomic alkali molecules that possess a permanent electric dipole moment in the ground electronic state. This attribute is of great interest in using ultracold molecules for quantum computation applications [8, 9], where controlling dipole-dipole interactions is key to using molecules as qubits.

One common method of creating ultracold molecules is photoassociation (PA). PA creates molecules in an electronically excited state, which quickly decay to high-lying vibrational levels of the lowest singlet (ground) and triplet electronic states. Transferring these molecules to the most deeply bound rovibrational

---

a) adeelmala@gmail.com

b) elliottd@purdue.edu

*Present address: Raman Research Institute, C. V. Raman Avenue, Sadashivanagar, Bangalore – 560080, India

ground-state levels, however, requires transfer pathways that may involve additional excitation steps to states that have a better wave function overlap with the rovibronic ground state [10, 11].

Our choice of molecule, $^7Li^{85}Rb$, has a large electric dipole moment (~4.1 Debye [12]) in the rovibrational ground state. Experimentally, LiRb is one of the least well-known heteronuclear alkali molecules that can be created in ultracold traps. Very recently, however, studies of two groups have started to shed light on this molecule through spectra in hot vapors [13, 14, 15], as well as, for the first time, PA spectroscopy using trap-loss measurements [16, 17]. Especially exciting is the high rate of photoassociation of this molecule, as reported in Refs. [16] and [17]. These studies have been guided by the results of several *ab initio* calculations [18, 19, 20].

In this paper, we report the formation of ultracold $^7Li^{85}Rb$ molecules in the lowest triplet state, $a^3\Sigma^+$, and perform preliminary spectroscopic studies of ultracold $^7Li^{85}Rb$ molecules using ionization detection. In the process, we have discovered new PA resonances that were not seen in earlier trap-loss spectra. In addition, we also measure resonantly-enhanced multiphoton ionization (REMPI) spectra, and identify vibrational levels of the following previously unobserved electronic states: $3^3\Pi$ ($v_\Pi' = 0 - 10$), $4^3\Sigma^+$ ($v_\Sigma' = 0 - 5$), and $a^3\Sigma^+$ ($v'' = 7 - 13$). We compare our observations with *ab initio* calculations [19] and semi-empirical potentials [13], and show good agreement between theory and experiment. These initial spectroscopic investigations shed new light on its structure, and are important steps toward finding transition pathways to transfer ultracold LiRb molecules into the most deeply bound rotational, vibrational and electronic (rovibronic) ground state levels.

The paper is organized as follows: In Section II we describe our experimental setup and detection scheme. We discuss PA spectroscopy in Section III, the REMPI spectra and their analysis in section IV, the estimation of molecule formation rate in section V and we conclude in Section VI.

## II. EXPERIMENTAL SETUP

In this section, we briefly describe the experimental parameters of our measurements. A more complete description can be found in Ref. [21]. We generate a sample of cold LiRb molecules by photoassociating $^7Li$ atoms and $^85Rb$ atoms in a dual species MOT. We show this PA step in Fig. 1(a), in which we display a diagram of the relevant potential energy curves (PECs) for LiRb [19]. The long-range PECs to which we photoassociate ultracold LiRb molecules are shown in Fig. 1(b). The excited state molecules decay spontaneously to the lowest triplet state, from which we ionize them using REMPI. Vibrational levels of the $3^3\Pi$ and $4^3\Sigma^+$ electronic states enhance the ionization signal when the ionizing laser is resonant with a transition from the $a^3\Sigma^+$ state. The sensitivity of this ionization scheme allows us to detect PA resonances that were unobservable through trap-loss spectroscopy, and the REMPI spectra allow us to determine energy levels of the ground and excited electronic states of the LiRb molecule that have not previously been observed.

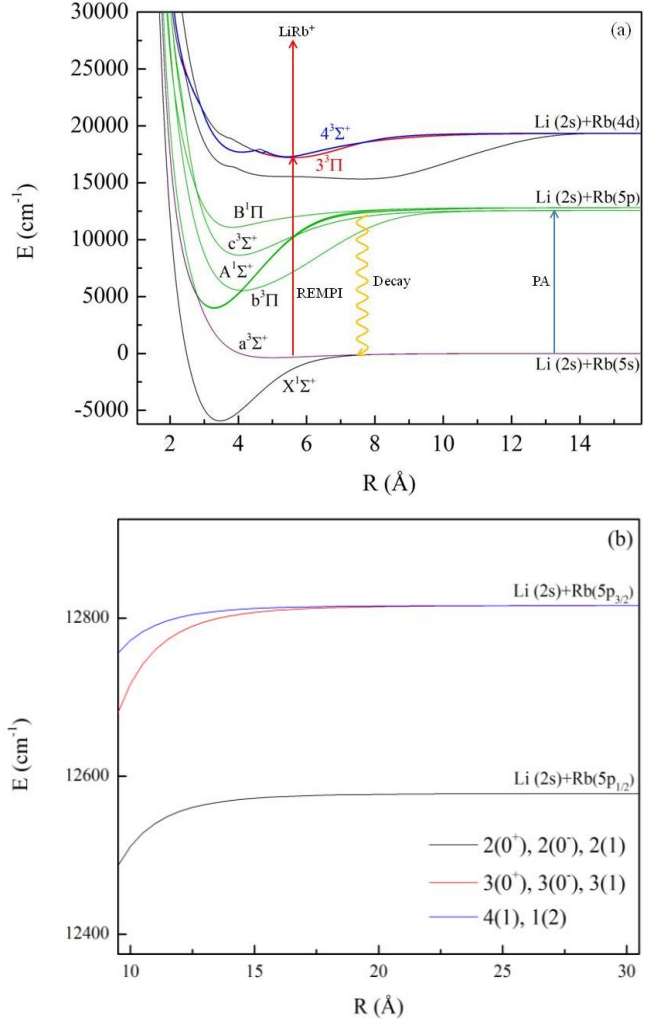

FIG. 1. (Color online) (a) Schematic of the PA and REMPI process, with selected potential energy curves relevant to this work plotted from *ab initio* calculations of Korek *et al.* [19]. PA molecules decay to the $a^3\Sigma^+$ state and are resonantly ionized by a pulsed laser through the $3^3\Pi$ or $4^3\Sigma^+$ intermediate electronic state. (b) Representation of the long range potentials asymptotic to the $^7$Li $2s$ $^2S_{1/2}$ + $^{85}$Rb $5p$ $^2P_{1/2}$ and $^7$Li $2s$ $^2S_{1/2}$ + $^{85}$Rb $5p$ $^2P_{3/2}$ free atomic states, using calculations of Ref. [22].

We cool and trap atomic $^7$Li and $^{85}$Rb in a dual species MOT, with atom number $N$ and density $n$ for Li and Rb as: $N_{Li} \sim 5 \times 10^7$, $n_{Li} \sim 5 \times 10^9$ cm$^{-3}$; $N_{Rb} \sim 1 \times 10^8$, $n_{Rb} \sim 4 \times 10^9$ cm$^{-3}$. We use a dark spot Rb MOT to reduce collisional losses in the trap [21], leaving most of the trapped $^{85}$Rb atoms in the $F = 2$ ground hyperfine level. We monitor the MOT operation and the overlap of the Li and Rb atomic clouds with a pair of CCD cameras. We photoassociate LiRb molecules using the 795 nm output of a Coherent 899 cw Ti:Sapphire (PA) laser, whose output power is typically between 300 and 350 mW and whose collimated beam diameter is ~0.85 mm. The frequency of the Ti:S laser beam is locked to a temperature stabilized external Fabry-Perot cavity, reducing the laser linewidth to ~1 MHz. We monitor the PA laser frequency with a Bristol 621A wavemeter (accuracy ~60 MHz). The excited state molecules created through PA spontaneously decay to weakly-bound vibrational levels of the $a^3\Sigma^+$ state, from which we ionize them using the output of a Spectra-Physics PDL-2 pulsed dye laser, pumped by the second harmonic of a Q-switched Nd:YAG (~7 ns pulse width, 10 Hz repetition rate). The PDL-2 output has a linewidth of ~0.3 cm$^{-1}$. We use Rhodamine 590 dye in the dye laser. For ionizing the molecules, we typically use a pulse energy of 2-3 mJ and a laser beam diameter of ~4 mm at the MOT. For REMPI spectroscopy, we fix the Ti:S laser frequency on a chosen PA resonance, and monitor the frequency drift by recording the heterodyne beat signal between part of the Ti:S output and a frequency comb (with a 250 MHz frequency difference between comb teeth). The linewidth of the beat signal is ~2 MHz. During normal operation, the beat signal may drift by ~5 MHz over a period of 30-45 minutes, which is the typical length of time required to collect a single data set. This drift is less than the typical linewidth of our PA resonances (~10s of MHz), and does not significantly affect our REMPI spectra.

Atomic and molecular ions created by the PDL pulses are accelerated towards a microchannel plate (MCP) detector assembly by a set of three field plates. We bias the MCP at a potential difference of -1.8 kV. Each ion strike registers as a sharp (~5 nsec) current spike, and we record these events to count the number of ions detected for each ionizing pulse. We use time-of-flight (TOF) mass spectrometry to resolve atomic and molecular ions, and to count only LiRb$^+$ ions. Our maximum count rate is ~2 LiRb$^+$ ions per laser pulse. For PA spectra, we fix the dye laser wavelength and scan the PA laser frequency, whereas for REMPI spectroscopy we fix the PA laser frequency and tune the output wavelength of the PDL. We control the Nd:YAG firing sequence, tune the dye laser wavelength and Ti:S laser frequency, and record the LiRb$^+$ ion counts using a laboratory PC.

## III. PHOTOASSOCIATION SPECTROSCOPY

We acquire PA spectra by scanning the PA laser in discrete ~20 MHz frequency steps below the $^7$Li 2s $^2$S$_{1/2}$ + $^{85}$Rb 5p $^2$P$_{1/2}$ asymptote. See Fig. 1(b) for the long-range PECs in this energy range. All the PA line positions $\Delta_{PA}$ in this section are measured relative to the $^{85}$Rb (5s, $F = 2$) − $^{85}$Rb (5p$_{1/2}$, $F = 2$) transition at $\nu_{1/2}$ = 377 108.946 GHz. We show the newly-observed PA resonances in Fig. 2. As in our trap-loss measurements [16, 17], we observe and assign resonances to all three electronic states [2(0$^+$), 2(0$^-$) and 2(1)] asymptotic to this level. For $\Delta_{PA}$ ~ 8 GHz or less, the PA laser disrupts the Rb MOT operation, rendering PA resonances unobservable in this range. As with trap-loss spectroscopy [16, 17], we only observe one rotational level for most of the PA resonances and assign these as $J = 1$. We include the REMPI wavelengths used to acquire these spectra in Fig. 2.

In our previous trap-loss spectroscopy work, we observed PA levels $u_{PA} = 2 - 4$ of $2(0^+)$ and $u_{PA} = 3 - 5$ of $2(0^-)$ and $2(1)$ (PA labeling starts at $u_{PA} = 1$ for the first PA level below the dissociation limit). In addition, we now also find lines for $u_{PA} = 5$ of $2(0^+)$, $u_{PA} = 8$, 9 and 11 of $2(0^-)$ and $u_{PA} = 6$, 7 and 10 of $2(1)$, visible only through ion detection. We include in Fig. 2 the PA resonances that we observed only using the REMPI detection scheme, but note that, for peaks that appear in both detection schemes, the line positions in trap-loss spectra are in excellent agreement (within ~40 MHz, consistent with the linewidth of the resonances) with the line positions that we now observe through ion detection. We also observe similar hyperfine structure in both sets of spectra. We have normalized all signals in Fig. 2 to the the $2(1)$ $u_{PA} = 5$ PA REMPI line strength.

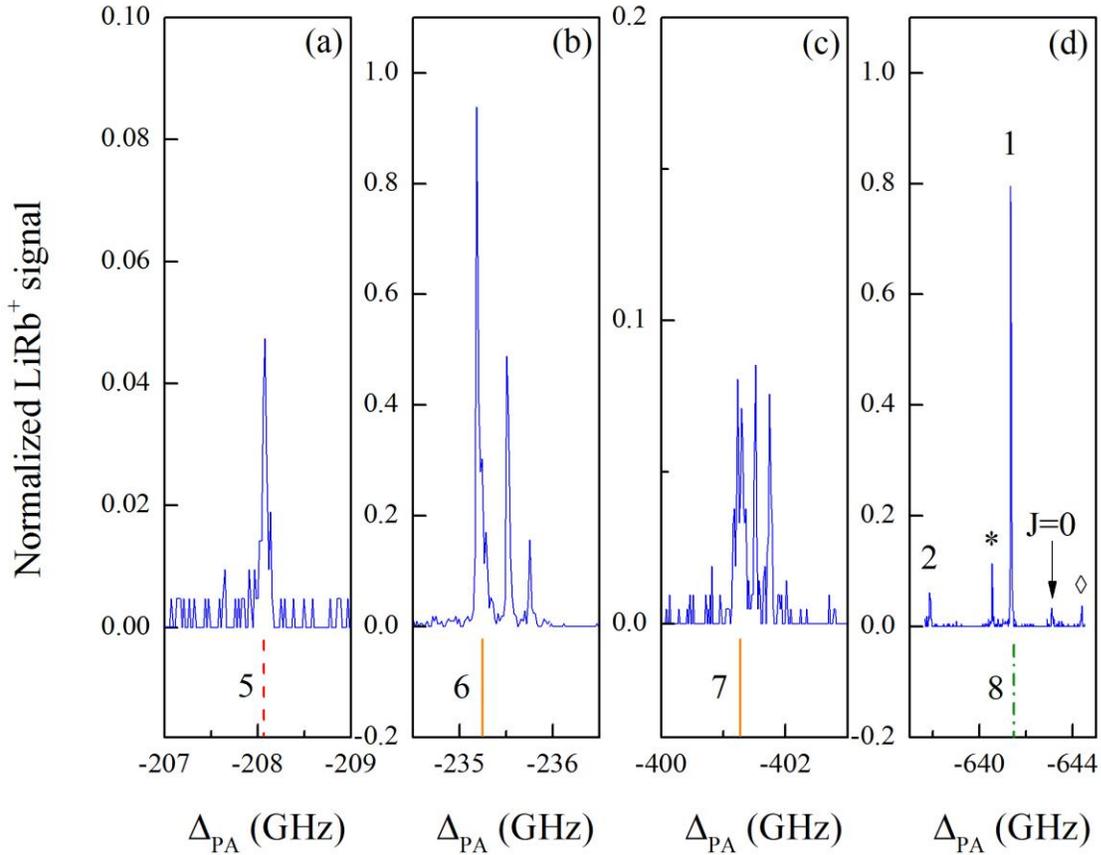

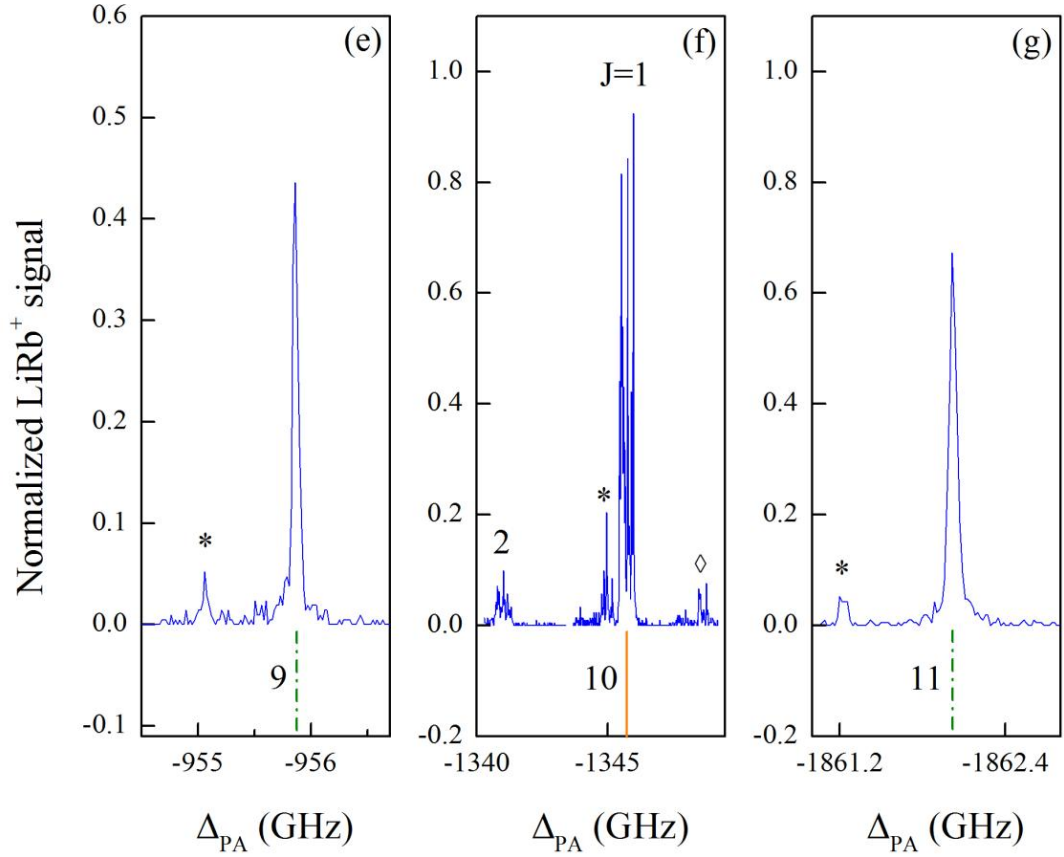

FIG. 2. (Color online) (a-g) The normalized PA resonances for the 2(0$^+$) (red dash), 2(0$^-$) (green dash-dot) and 2(1) (orange solid) states. We have labeled each PA vibrational level $u_{PA}$ below the respective resonance. We identify the 2(1) resonances [in panels (b), (c) and (f)] through their hyperfine structure. By contrast, the $\Omega = 0$ lines show no hyperfine structure, except for loosely bound levels (2(0$^-$) $u_{PA} = 5$, for example). We have labeled multiple rotational levels where observed (panels (d) and (f)); otherwise the rotational quantum number is J=1. The asterisks (*) and diamonds ($\diamond$) label hyperfine ghosts originating from population in the Li(2s, $F = 1$) and Rb(5s, $F = 3$) states, respectively. The ionization laser frequencies are: (a) − (c) 17742.4 cm$^{-1}$, (d) − (f) 17602.5 cm$^{-1}$ and (g) 17612.7 cm$^{-1}$.

**Peak assignment**

As done in our previous reports [16, 17], we base electronic state assignments of the PA levels on the structure of the PA resonances. The hyperfine energy, $E_{hfs}$, scales as $E_{hfs} \sim \Omega/J(J+1)$ [23], where $\Omega$ is the projection of the total angular momentum, excluding nuclear spin, on the internuclear axis. Hence, $\Omega = 0$ lines have no internal structure to first order but $\Omega \geq 1$ lines show hyperfine structure. In Fig. 2 we show the sets of peaks assigned to each of the three electronic states.

In Table I, we list the observed PA lines for the $J = 1$ line of each vibrational level. Through earlier trap-loss measurements, we identified PA lines up to $u_{PA} = 4$ for $2(0^+)$, and $u_{PA} = 5$ for $2(0^-)$ and $2(1)$ [17]. With the more sensitive ionization detection method, we identify an additional seven PA levels, as listed below the dashed lines in Table I, with a binding energy of up to $\Delta_{PA} \sim -62$ cm$^{-1}$ ($\sim$1862 GHz below the $^7$Li 2s $^2$S$_{1/2}$ + $^{85}$Rb 5p $^2$P$_{1/2}$ asymptote. One additional weak PA resonance $\sim$218 GHz below the $^7$Li 2s $^2$S$_{1/2}$ + $^{85}$Rb 5p $^2$P$_{1/2}$ resonance that was unobserved with earlier trap-loss spectroscopy is also shown in the supplementary material [24]). We use some of these newly discovered PA lines to record REMPI spectroscopy of ground and excited electronic states (next section).

In Ref. [17], we derived the $C_6$ dispersion coefficients of the PA states by fitting the binding energy of the three vibrational levels observed for each electronic series through trap-loss spectra to the Le-Roy Bernstein formula [25] for a long range potential described by a single $C_6/R^6$ term. It becomes of interest now to re-compute $C_6$ using our extended data sets for each of these series. We find that the potential is well characterized by $C_6/R^6$ over these extended energy ranges. The revised values of $C_6$ are in good agreement with the original values ($C_6 = 10120$ (450) for the $2(0^+)$, 12410 (180) for the $2(0^-)$, and 13130 (330) for the $2(1)$ electronic states).

We note that we searched for, but were unable to observe, the PA resonances that are missing in Table I, such as $u_{PA} = 6$, 7 and 10 of the $2(0^-)$ state, and $u_{PA} = 8$ and 9 of the $2(1)$ state. Similarly, we were unable to observe more deeply bound PA levels than those shown in Table I.

Table I. Binding energies $\Delta_{PA}$ (in GHz) for the observed PA levels below the Li(2s) + Rb(5p$_{1/2}$) asymptote at $v_{1/2}$ = 377108.946 GHz. States below the dashed line in each column are new, observed through the REMPI detection scheme alone. The three 2(0⁻) state entries shown in boldface indicate the PA resonances used for the REMPI spectra discussed in Section IV. Note: 29.9792458 GHz = 1 cm⁻¹.

| $u_{\mathrm{PA}}$ | 2(0⁺) | 2(0⁻) | 2(1) |
|---|---|---|---|
| 2 | -9.45 | | |
| 3 | -38.06 | -15.48 | -16.26 |
| 4 | -98.44 | -51.04 | -52.54 |
| 5 | -208.08 | **-119.21** | -121.41 |
| 6 | | | -235.51 |
| 7 | | | -401.52 |
| 8 | | -641.33 | |
| 9 | | **-955.86** | |
| 10 | | | -1345.77 |
| 11 | | **-1862.02** | |

## IV. SPECTROSCOPY OF $a^3\Sigma^+$, $3^3\Pi$ AND $4^3\Sigma^+$ STATES

As mentioned earlier, the electronic ground state molecules formed by spontaneous decay of photoassociated molecules can be ionized via two-photon absorption using the output pulses of the PDL, when the frequency of this laser is one-photon resonant with a transition between a populated ground state, with vibrational level $v''$, and an excited intermediate state, with vibrational level $v'$. We show a schematic representation of this REMPI process in Fig. 1. The REMPI spectra allow us to identify the energies of the lowest triplet $a^3\Sigma^+$ state and the intermediate $3^3\Pi$ and $4^3\Sigma^+$ states.

We acquire the spectra presented in this section by fixing the PA laser frequency at one of the three different PA resonances: $u_{\mathrm{PA}}$ = 5, 9 and 11, all belonging to the 2(0⁻) electronic state, and tuning the output frequency of the PDL in discrete ~ 0.3 cm⁻¹ steps. We record spectra for $u_{\mathrm{PA}}$ = 5 and 11 for PDL frequencies between 17280 cm⁻¹ and 18050 cm⁻¹ (wavelengths ~578.7 and 554.0 nm, respectively), and between 17589 cm⁻¹ (~568.5 nm) and 17821 cm⁻¹ (~561.1 nm) for $u_{\mathrm{PA}}$ = 9. From one-photon electric-dipole selection rules for transitions valid for Hund's case c [$\Delta\Omega$ = 0, ±1 except 0⁺ ↔ 0⁻], molecules from the 2(0⁻) state decay to 1(0⁻) and 1(1) ground states, which correlate to the $a^3\Sigma^+$ state. For the stated PDL wavelength range, LiRb molecules ionize through intermediate electronic states that are asymptotic to the Li(2s) + Rb(4d$_j$) states ($j$ = 3/2 and 5/2). In the experimental line assignments presented below, we estimate the uncertainty in the transition energies to be ~0.5 cm$^{-1}$, mainly due to the PDL linewidth and shot-to-shot fluctuations of the laser frequency.

**Analysis of spectra**

In Fig. 3a, we show the complete REMPI spectrum between 17280 and 18050 cm$^{-1}$, recorded with the PA laser fixed at the $2(0^-)$ $u_{PA} = 11$. As we will show in this section, the frequencies of these spectral lines, as well as those of similar spectra that we recorded with the PA laser fixed at the $u_{PA} = 5$ or $u_{PA} = 9$ lines of the $2(0^-)$ state, allow us to identify the relative energies of several vibrational levels of the lowest triplet $a^3\Sigma^+$ state, the excited $3^3\Pi$ state, and the excited $4^3\Sigma^+$ state.

**Features associated with $a^3\Sigma^+$ $v''$ – $3^3\Pi_\Omega$ $v_\Pi'$ transitions**

The REMPI spectra reveal structure on three different energy scales. On the largest scale, we observe recurring features spaced by ~70 cm$^{-1}$, as are clearly seen in the total spectrum of Fig. 3a. In Fig. 3b, we show a section of the REMPI spectrum between 17725 and 17825 cm$^{-1}$ to highlight some of the identified peaks and the distinct features seen in the spectra (the complete labeled spectra are available in the Supplementary information [24]). In this figure, these features are seen as a grouping of three peaks around 17740 cm$^{-1}$ and around 17810 cm$^{-1}$. We assign these features separated by ~70 cm$^{-1}$ as the progression of $3^3\Pi_\Omega$ vibrational levels $v'_\Pi$, where we include the spin-orbit component of this state as a subscript: $3^3\Pi_\Omega$.

Each $v'_\Pi$ line also has a characteristic triplet substructure, with adjacent 'teeth' within each triplet regularly spaced by ~3 cm$^{-1}$. We observe that each tooth within a triplet shows almost equal signal strength. This is seen in each of the three sets of $u_{PA}$ REMPI spectra. We assign these triplets – from lowest to highest frequency – as the $\Omega = 0$, 1 and 2 spin-orbit components of $3^3\Pi$, as we have indicated on just one of these triplets near 17775 cm$^{-1}$ in Fig. 3b.

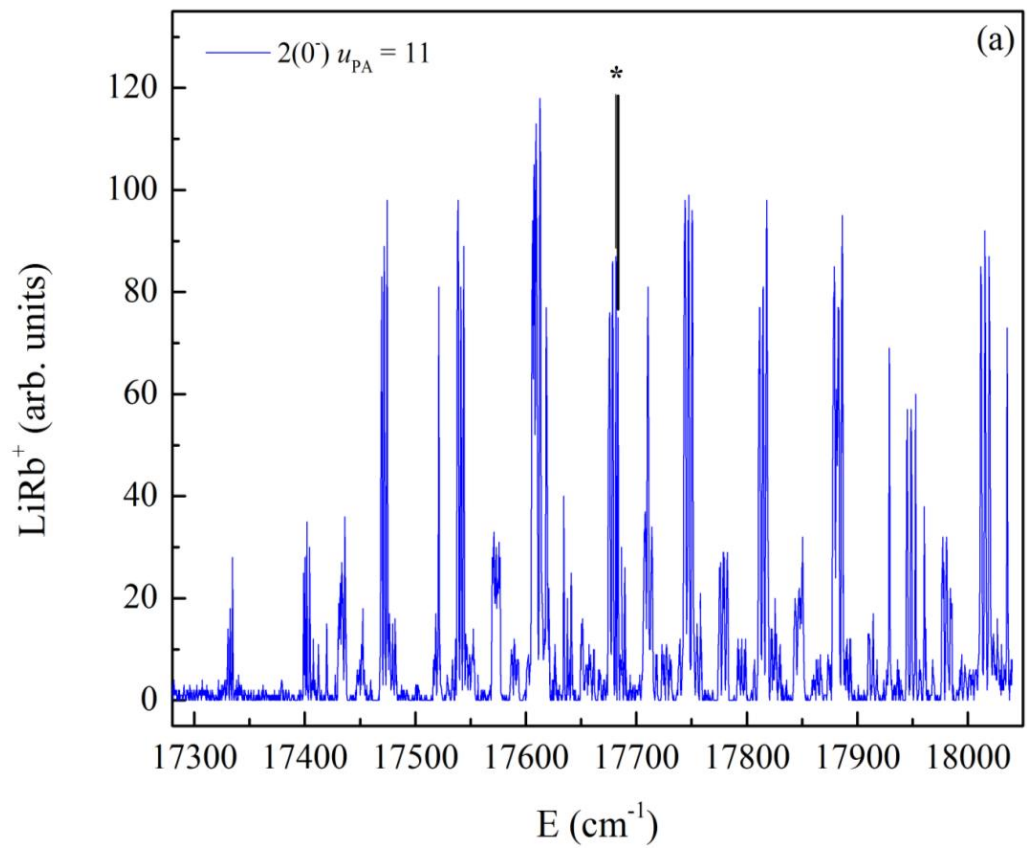

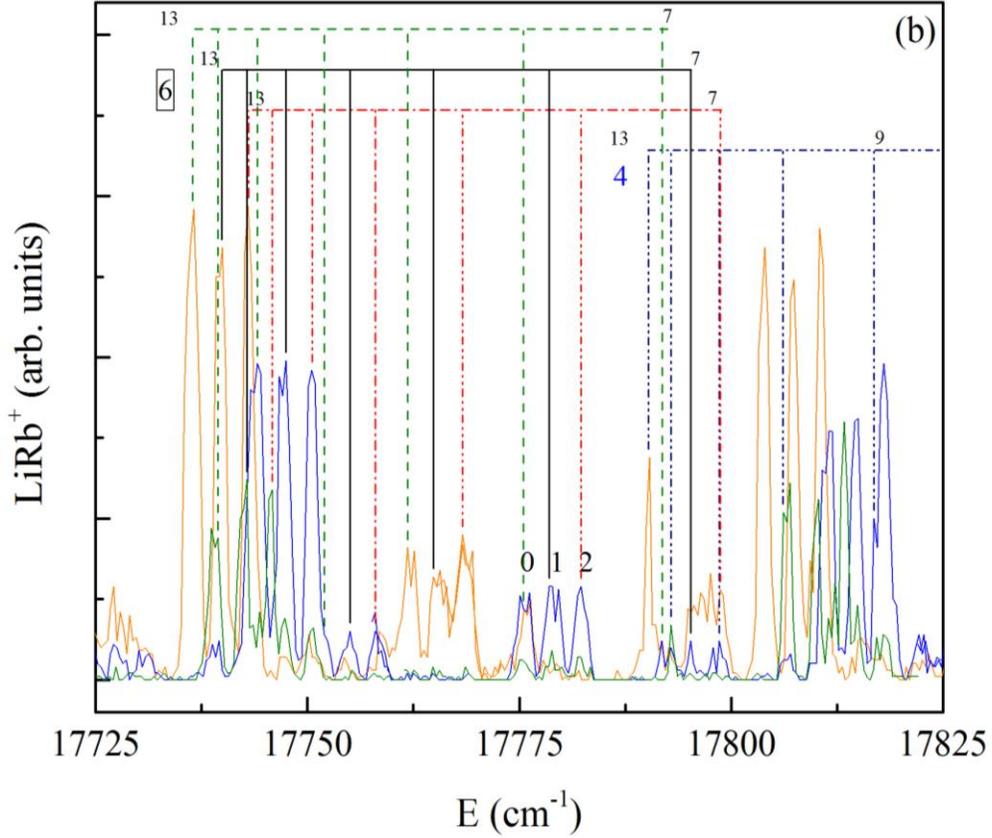

FIG. 3. (Color online) (a) REMPI spectrum recorded for the $2(0^-)$ $v = 11$ PA resonance. The asterisk (*) denotes the Rb 9s $^2S_{1/2}$ - 5p $^2P_{3/2}$ one-photon transition, and the 9s $^2S_{1/2}$ - 5p $^2P_{3/2} - 5s$ $^2S_{1/2}$ stepwise transition due to the presence of the Rb MOT pump laser (5p $^2P_{3/2} - 5s$ $^2S_{1/2}$) and the REMPI ionizing laser (9s $^2S_{1/2}$ - 5p $^2P_{3/2}$). (b) A short representative section of the REMPI spectrum, showing features and patterns originating from $3^3\Pi$ $v'_\Pi$, $4^3\Sigma^+$ $v'_\Sigma$ and $a^3\Sigma^+$ $v''$. The spectra originate from $u_{PA}$ = 5 (orange), 9 (green) and 11 (blue), respectively. Numbers on top of the horizontal lines represent $v''$ of the $a^3\Sigma^+$ state, where the green dashed, black solid and red dash-dot lines label $3^3\Pi$ $\Omega$ = 0, 1 and 2 components, respectively (labeled for the set of triplets near 17775 cm$^{-1}$). The boxed number to the left of the (green) dashed vertical line denotes the vibrational number $v_\Pi'$ of the $3^3\Pi_\Omega$ state. The unboxed number to the left of the (blue) dash-dot-dot vertical line labels $v_\Sigma'$ of the $4^3\Sigma^+$ state. See supplementary material [24] for the complete set of REMPI spectra with the line positions and their assignments.

Finally, comparing the overlayed REMPI spectra of $u_{PA}$ = 5, 9 and 11 reveals several different $a^3\Sigma^+$ vibrational levels $v''$. This is the progression with increasing spacing of the triplet set of peaks as we go towards higher frequency. Transitions from high-lying vibrational levels $v''$ near the dissociation limit appear stronger than transitions from more deeply bound (lower $v''$) levels.

Having identified line positions and features corresponding to $3^3\Pi_\Omega$ $v_\Pi' \leftarrow a^3\Sigma^+$ $v''$ transitions, we now present details of vibrational level assignment of the $3^3\Pi_\Omega$ states.

### $3^3\Pi_\Omega$ vibrational levels and spin-orbit components

In order to assign the line positions to vibrational levels of an excited electronic state, we make use of the results of the *ab initio* calculations of LiRb PECs by Korek *et al.* [19], which include spin-orbit interaction. We employ the potential curve of the $a^3\Sigma^+$ state calculated by Ivanova *et al.* [13] for calculations of the energy spacing between vibrational levels $v''$ of the ground state, and LEVEL 8.0 [26] to calculate electronic state vibrational levels and Franck-Condon factors.

The *ab initio* potentials with spin-orbit interactions show numerous avoided crossings. For these adiabatic potentials, we assume separation of nuclear and electronic motion based on the Born-Oppenheimer approximation, which gives rise to avoided crossings between electronic states of the same symmetry. However, the relativistic wavefunctions in the vicinity of these avoided crossings change rapidly and result in non-adiabatic effects [27]. Hence, we modify the *ab initio* PECs asymptotic to Li(2s) + Rb(4d$_j$) and Li(2s) + Rb(6s) by using diabatic crossings and recalculating the PECs. We show the modified diabatic PECs in Fig. 4(a), where we highlight the relevant $3^3\Pi$ and $4^3\Sigma^+$ potentials.

The $3^3\Pi$ state consists of four spin-orbit components: $\Omega = 0^-, 0^+, 1$ and 2. From LEVEL 8.0 calculations of the diabatic PECs, we calculate and show the spin-orbit components in Fig. 4. At the equilibrium internuclear separation $R_e$ ~5.6 Å, the calculated energy splittings $\Delta E_\Omega$ between the various $\Omega$ components are unequal. Our REMPI spectra, however, reveal a slightly different picture: we observe only three out of four spin-orbit components, and the spacing $\Delta E_\Omega$ between the *components* of the triplet structures appears to be nearly equal. While we use the diabatic PECs from Korek *et al.* [19] to aid our assignment of the spin-orbit components (see Figure 3b), the qualitative differences between calculated and observed levels makes these assignments tentative. We hope that our measurements will motivate development of an improved model for the $3^3\Pi$ state. We note that degeneracy of the $0^+$ and $0^-$ components, as well as roughly equal spacing between $\Omega = 0$ and 1, and $\Omega = 1$ and 2 has also been observed in REMPI spectroscopy performed in KRb [28].

Next, we assign the vibrational levels $v_\Pi'$ by comparing calculated vibrational levels and spacing with our experimental results. We tentatively assign $v_\Pi'$ starting with 0 because in our REMPI spectra we do not observe resonances at frequencies below ~17323 cm⁻¹. While this could be due to diminishing Franck-Condon factors or the limited tuning range of the Rhodamine 590 dye (or both), assigning these lowest observed lines as $v_\Pi{}' = 0$ is supported by the ~89.4 cm⁻¹ energy spacing between the lowest vibrational state of the $3^3\Pi_{\Omega = 0}$ and $4^3\Sigma^+$ states, in good agreement with the ~85.5 cm⁻¹ spacing predicted in the theoretical PECs of Ref. [19]. (We discuss the assignment of the $4^3\Sigma^+$ vibrational levels in the next section.)

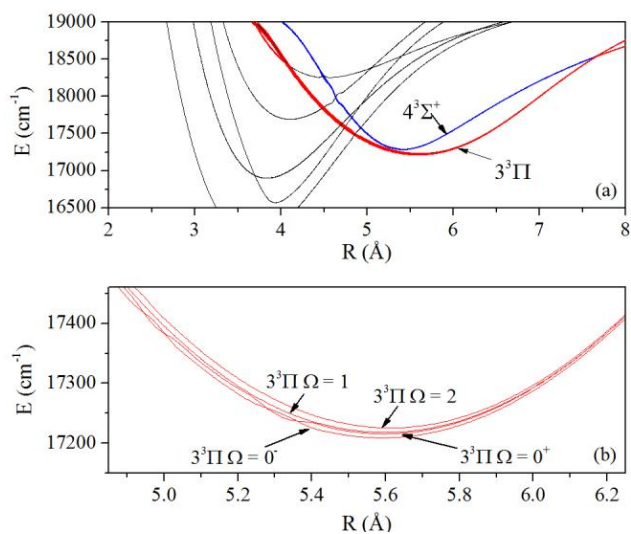

FIG. 4. (Color online) (a) Plots of the diabatic PECs. In our studies, we identify excited state vibrational level progressions associated with the $3^3\Pi$ (red) and $4^3\Sigma^+$ states (blue). (b) Close-up view of the $3^3\Pi$ state showing the spin-orbit splitting. All curves are plotted from *ab initio* calculations of Korek *et al.* [19]

We also use the calculated vibrational spacings $\Delta G_{th}$ of Ref. [19] to aid in assigning the vibrational levels of the $3^3\Pi$ state. We show these data in Fig. 5, in which the open red circles show the calculated level spacings, and the black closed dots show the experiment spacings. We observe very good agreement between theory and experiment for these data.

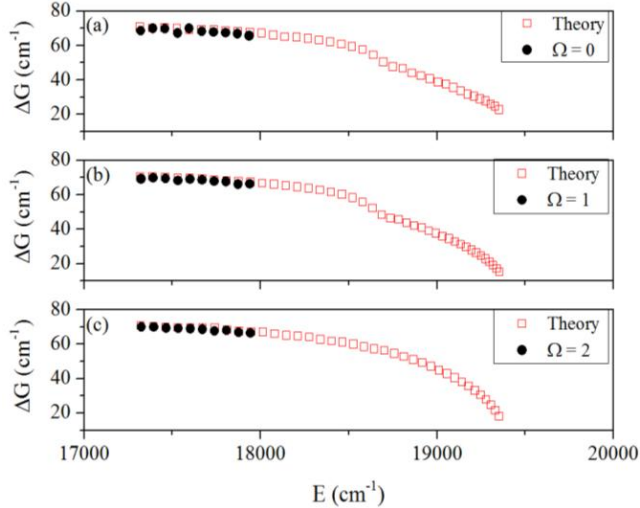

FIG. 5. (Color online) Comparison between calculated [19] and observed $3^3\Pi$ vibrational spacings $\Delta G_{v''+1/2}$ for (a) $\Omega = 0$, (b) $\Omega = 1$ and (c) $\Omega = 2$. Theoretical calculations (red open squares) are based on the diabatic PECs and calculated using LEVEL 8.0 [26]. The experimental data points (black closed circles) are each plots of $\Delta G_{exp}$ for $\Omega = 0$, 1 and 2.

### $a^3\Sigma^+$ vibrational structure

We assign $a^3\Sigma^+$ $v''$ levels by examining line positions in the $u_{PA} = 5$, 9 and 11 REMPI spectra and comparing our observations with the vibrational level spacing $\Delta G_{v''+1/2}$ calculated from the analytic $a^3\Sigma^+$ PEC of Ivanova *et al.* [13]. Experimental $\Delta G_{v''+1/2}$ most closely match calculated $\Delta G_{v''+1/2}$ for $v'' = 7 - 13$. In Table II we compare the calculated vibrational energy spacing $\Delta G_{v''+1/2}$ with experiment. Entries in the column titled $\Delta G_{v''+1/2}$ (Exp.) are the average of the observed vibrational spacing for all three $3^3\Pi_\Omega$ spin-orbit components, which we derive from the present spectra. Although the agreement is very good, one should accept these assignments as tentative, since the PEC of Ref. [13] is based on a very limited set of spectral lines.

We observe (Fig. 3b) that transitions from $v'' = 13$, 12 and 11 are strongest when the PA levels used are $u_{PA} = 5$, 9 and 11, respectively, *i.e.* spontaneous decay from more deeply bound PA resonances leads to population in more deeply bound vibrational levels of the $a^3\Sigma^+$ state. This strongly suggests the underlying reason for the different line strengths to be the varying wavefunction overlap between the $u_{PA}$ levels and $v''$. The strength of these $u_{PA} - v''$ transitions, in part, can be qualitatively explained by Franck-Condon factors (FCF) for decay from the initial PA vibrational levels to the ground triplet state $v''$. We calculate the FCFs from $u_{PA} = 5$, 9 and 11 to $a^3\Sigma^+$ $v''$ using the $2(0^-)$ *ab initio* potential from Korek *et al.* and the $a^3\Sigma^+$ potential from Ivanova *et al.*, and show these in Fig. 6. We see that the FCFs for deeper PA levels are larger for more deeply bound ground state vibrational levels, in qualitative agreement with our observations. The calculated FCFs differ in detail from our observations, however, in the distribution of $v''$ populated through each PA resonance. With the PA laser fixed on $u_{PA} = 5$, we primarily populate $v'' = 13$, 9 and 7 of the ground state; $u_{PA} = 9$ populates $v'' = 12$ and 8; and $u_{PA} = 11$ populates $v'' = 11$, 10, 8 and 7. For the $v''=13$-11 vibrational levels of the $a^3\Sigma^+$ state, the agreement with the calculated FCFs is reasonable. For the $v''=7$-10 vibrational levels, however, the theoretical predictions for the transitions from the PA levels do not agree with the experimental observations of the REMPI spectra. Several factors could play a role in such disagreement: the accuracy of the excited state, the presence of avoided crossings among different excited states and the ionization probability. The excited state used for the FCF calculation is based on fully *ab initio* calculations, and taking into account the sensitivity of the wave function to small changes in the interaction, a difference between calculations and experiment would be expected. Our calculations are based on a single channel approach, where the excited states are assumed to be uncoupled. This approximation is less accurate if the excited states show interaction among them through the existence of avoided crossings. Finally, we analyze the effect of the ionization probability, since deeper bound states show higher ionization probability than shallow ones. This would explain qualitatively why deeper vibrational levels of the $a^3\Sigma^+$ state show more intense REMPI lines than is predicted from FCFs calculation. It is worth mentioning that similar disagreement between calculations and experiment has been observed in other REMPI spectra [26, 28]. Revisiting the excited state PECs with the help of experimental data might help to improve theoretical calculations, which in turn may lead to a more accurate calculation of the FCFs.

Additional observations from our spectra are that: (*i*) the $-3^3\Pi$ $v_\Pi' \leftarrow a^3\Sigma^+$ ($v'' = 7$) features are weaker than the abovementioned transitions for all three spectra, likely due to only minimal decay to this vibrational level from any of the PA resonances used in our measurements; and (*ii*) many $v'' = 10$ transitions are partially obscured by the overlap with stronger $v'' = 11$ lines. These transitions are most clearly seen in the 17440 – 17600 cm$^{-1}$ range and labeled in Figure S5(b) in the Supplementary material [24].

Table II. $a^3\Sigma^+$ vibrational levels and energies (measured from the Li(2s) + Rb(5s) asymptote: $E_d$ calculated from the analytic PEC of [13]). The calculated vibrational spacing ($\Delta G_{v''+1/2}$ (Th.)) is compared with experimental results ($\Delta G_{v''+1/2}$ (Exp.)) and shows very good agreement. The ($E_{v''} - E_d$) column lists the calculated vibrational level energies $E_{v''}$ with respect to the dissociation energy $E_d$. For $a^3\Sigma^+$ $v''$ identification, the vibrational level spacing $\Delta G$ is more relevant than the calculated $E_{v''}$ values. All energies and vibrational spacings are in units of cm$^{-1}$.

| $v''$ | $(E_{v''} - E_d)$ | $\Delta G_{v''+1/2}$ (Th.) | $\Delta G_{v''+1/2}$ (Exp.) | $\Delta G_{v''+1/2}$ (Th.) $-$ $\Delta G_{v''+1/2}$ (Exp.) |
|---|---|---|---|---|
| 7 | -57.0 | 16.7 | 16.6 | 0.1 |
| 8 | -40.3 | 13.6 | 13.7 | -0.1 |
| 9 | -26.7 | 10.4 | 10.4 | 0.0 |
| 10 | -16.3 | 7.5 | 7.7 | -0.2 |
| 11 | -8.8 | 5.0 | 4.7 | 0.3 |
| 12 | -3.8 | 2.6 | 2.8 | -0.2 |
| 13 | -1.1 | 1.0 | -- | -- |
| 14 | -0.1 | -- | -- | -- |

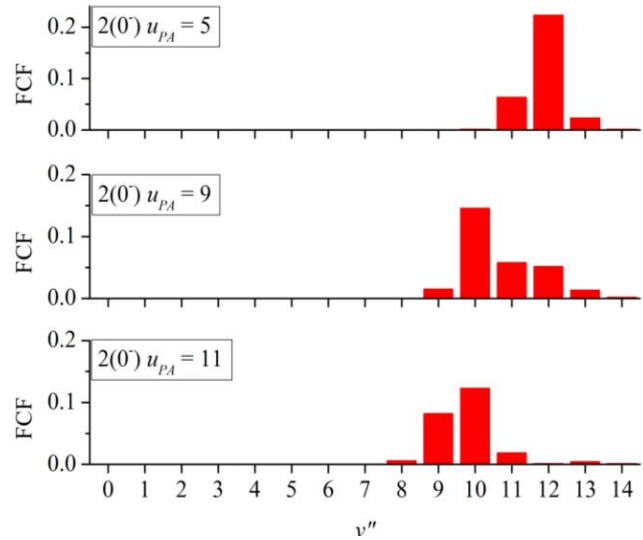

FIG. 6. (Color online) Calculated Franck-Condon factors (FCF) for transition from the three PA levels to $a^3\Sigma^+$ vibrational levels. For PA resonances with large binding energies, the FCF is larger for decay to lower (deeper bound) $v''$ levels.

**Assignment of $4^3\Sigma^+$ vibrational levels**

In addition to the $3^3\Pi_\Omega$ vibrational levels, we also observe and assign $v_\Sigma' = 0 - 5$ vibrational levels of the $4^3\Sigma^+$ state. The $4^3\Sigma^+$ diabatic PEC is shown as the blue curve in Fig. 4(a). Unlike the triplet structure originating from spin-orbit splitting of the $3^3\Pi$ state, we do not observe separate $\Omega = 0^-$ and 1 components for $4^3\Sigma^+$. This may be due to the small spin-orbit splitting for this state, calculated to be less than 1 cm$^{-1}$ at the equilibrium internuclear spacing $R_e \sim 5.4$ Å. We compare the calculated $4^3\Sigma^+$ vibrational spacing with experiment graphically in Fig. 7.

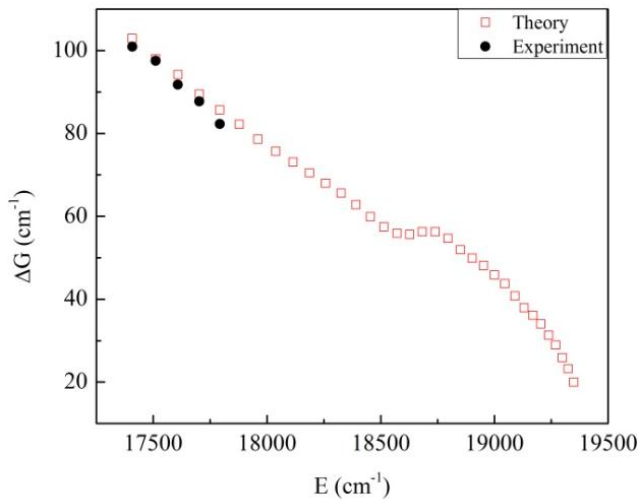

FIG. 7. (Color online) Comparison between calculated [19] and observed $4^3\Sigma^+$ vibrational energy spacings $\Delta G_{v'+1/2}$. Theoretical calculations (red open squares) are based on the diabatic PECs and calculated using LEVEL 8.0 [26]. The experimental data points are shown as black closed circles.

We recalculate the $v'' - v_\Pi'$ transition energies with reference to the Li(2s) + Rb(5s) asymptote for each spin-orbit component $\Omega = 0$, 1 and 2 of the $3^3\Pi$ and $4^3\Sigma^+$ states, and present these data in Table III. The calculated energies [19] of these states fall approximately 62-66 cm$^{-1}$ lower than the observed energies, indicating that the depth of the *ab initio* potential energy well is too large by this amount. This seems a very reasonable agreement with experiment.

Table III. Energies (in cm$^{-1}$) of the vibrational levels of the $3^3\Pi$ and $4^3\Sigma^+$ states as determined from the observed spectra. All energies are referenced to the Li(2s) + Rb (5s) asymptote.

| $v'$ | $3^3\Pi$ | | | $4^3\Sigma^+$ |
|---|---|---|---|---|
| | $\Omega = 0$ | $\Omega = 1$ | $\Omega = 2$ | |
| 0 | 17321.9 | 17324.2 | 17326.1 | 17411.8 |
| 1 | 17390.4 | 17393.4 | 17396.0 | 17512.7 |
| 2 | 17460.3 | 17463.2 | 17465.9 | 17610.3 |
| 3 | 17530.2 | 17532.7 | 17535.2 | 17702.0 |
| 4 | 17597.3 | 17601.0 | 17604.3 | 17789.8 |
| 5 | 17667.3 | 17670.0 | 17673.2 | 17872.0 |
| 6 | 17735.4 | 17738.6 | 17741.8 | |
| 7 | 17803.0 | 17806.5 | 17809.5 | |
| 8 | 17870.2 | 17874.1 | 17877.4 | |
| 9 | 17936.9 | 17940.2 | 17944.2 | |
| 10 | 18002.6 | 18006.6 | 18010.6 | |

We have fit the $3^3\Pi$ and $4^3\Sigma^+$ electronic state energies presented in Table IV to

$$E_v = T_e + \omega_e (v + 1/2) - \omega_e x_e (v + 1/2)^2 \qquad (1)$$

where $T_e$ is the term value, $\omega_e$ the vibrational constant, and $\omega_e x_e$ the anharmonicity, in order to determine the best values for each of these constants.

Table IV. The term value $T_e$, the vibrational constant $\omega_e$, and the anharmonicity $\omega_e x_e$ of the $3^3\Pi$ and $4^3\Sigma^+$ states as determined from the observed spectra. All energies are given in cm$^{-1}$, and are referenced to the Li(2s) + Rb (5s) asymptote. For comparison, we also list theoretical values, when available, from Refs. [19, 20]. [a]This work does not include spin-orbit coupling. These spectroscopic constants are expected to correspond most closely with $\Omega=1$ of the $3^3\Pi$ state, and $\Omega=0^-$ of the **$4^3\Sigma^+$** state. [b]We do not resolve $\Omega=0^+$ and $\Omega=0^-$ states in the present work. [c]We do not resolve $\Omega=0^-$ and $\Omega=1$ states in the present work. [d]Data given is for the second minimum of the potential, relevant to the present measurements. The values for the first minimum of this potential are $T_e$ = 17,372 cm$^{-1}$, $\omega_e$ = 117.85 cm$^{-1}$, and $\omega_e x_e$=1.74 cm$^{-1}$.

| State | | This work | Ref. [20][a] | Ref. [19] | |
|---|---|---|---|---|---|
| | | | | $\underline{\Omega = 0^+}$ | $\underline{\Omega = 0^-}$ |
| $3^3\Pi$ | $T_e$, | 17250.9 (7) | -- | 17224 | -- |
| $\Omega = 0$[b] | $\omega_e$ | 70.3 (3) | -- | 70.91 | -- |
| | $\omega_e x_e$ | 0.19 (3) | -- | -- | -- |
| $3^3\Pi$ | $T_e$, | 17253.3 (5) | 17301 | -- | |
| $\Omega = 1$ | $\omega_e$ | 70.5 (2) | 74.29 | -- | |
| | $\omega_e x_e$ | 0.19 (2) | 1.09 | -- | |
| $3^3\Pi$ | $T_e$, | 17255.3 (3) | -- | 17234 | |
| $\Omega = 2$ | $\omega_e$ | 70.6 (1) | -- | 70.86 | |
| | $\omega_e x_e$ | 0.19 (1) | -- | -- | |
| | | | | $\underline{\Omega = 0^-}$ | $\underline{\Omega = 1}$ |
| | $T_e$, | 17306.4 (8) | 17372[d] | 17699 | 17700 |
| $4^3\Sigma^+$ [c] | $\omega_e$ | 106.3 (6) | 109.58[d] | 108.26 | 118.31 |
| | $\omega_e x_e$ | 2.4 (1) | 1.74 [d] | -- | -- |

The term values are referenced to the Li(2s) + Rb (5s) asymptote, and, for comparison, we also list theoretical values from Refs. [19, 20], when available. We note good agreement between our measured values and the theory values in most cases, with the exception of the anharmonicity of the $3^3\Pi$ states, for which the theory value is five times greater than the measured value.

## V. MOLECULE FORMATION RATE

Before we conclude, we present an estimate of the number of triplet ground state molecules available for REMPI ionization. In earlier trap-loss studies [17], we observed an excited state molecule formation rate, $R_{PA}$, of ~$10^7$ s$^{-1}$ for resonances near the $^7$Li 2s $^2$S$_{1/2}$ + $^{85}$Rb 5p $^2$P$_{1/2}$ asymptote. Here, we use ionization data for the 2(0$^-$) $u_{PA}$ = 5 level to estimate $R_{PA}$. The following calculation, based on Ref. [29], takes into account the molecular creation rate by PA, ionization and detection efficiencies that will be relevant to ionization detection of ground state molecules as well.

The number of ions detected per ionization pulse, $N_{ion}$, depends on i) $N_a$, the number of ground (triplet) state molecules in a particular vibrational state (for example, v"=12), ii) the probability, $p_{ion}$, of ionizing molecules per ion pulse, and iii) the detection efficiency $e_d$. In other words,

$$N_{ion} = N_a p_{ion} e_d \qquad (2)$$

For the 2(0$^-$) $u_{PA}$ = 11 level, we record a strong ion signal $N_{ion}$ ~ 2 ions/pulse. The ionization probability is calculated as:

$$p_{ion} = 1 - e^{-\sigma E \lambda / hc\pi\omega^2}, \qquad (3)$$

where $\sigma$ is the photoionization cross-section, and since there is no available data for LiRb in the literature, we estimate it as the same order of magnitude as Rb$_2$ ~ $10^{-22}$ m$^2$ [29]. With pulse energy $E$ ~ 3 mJ, ionizing laser wavelength $\lambda$ = 565 nm and the pulsed-laser beam diameter, $2\omega$ = 4 mm, we obtain from Eq. (3) $p_{ion}$ ~ $6.5 \times 10^{-2}$. Estimating the detection efficiency $e_d$ is difficult, as it is related to the MCP detector efficiency and also the probability that an ion generated by the ionization laser strikes the MCP. Here, we estimate $e_d$ between 0.1 and 0.8. Using the values for $p_{ion}$ and $e_d$ along with $N_{ion}$, we calculate from Eq. (2) an estimate for the number of molecules in the $a^3\Sigma^+$ state: $N_a$ ~ 40 – 300/pulse. *Note:* From this, it is possible to estimate the number of $a^3\Sigma^+$ state molecules formed per second. Since molecules are not trapped they drift out of the ionization region in ~10 ms, while the ionizing laser pulse arrives only once every 100 ms allowing only ~10% of molecules to be detected. The number of $a^3\Sigma^+$ state molecules formed in 100 ms is thus ~10 $N_a$ i.e. the $a^3\Sigma^+$ state molecule formation rate is ~ 4x10$^3$ – 3x10$^4$ s$^{-1}$, relatively high and consistent with our previous expectations [16, 17].

Solving the rate equation for the number of $a^3\Sigma^+$ state molecules, a relation between the photoassociation rate $R_{PA}$, and the number of those molecules is found [29]:

$$N_a = \frac{R_{PA}(FCF)t}{1 + t/\tau} \qquad (4)$$

where $t$ is the time between REMPI laser pulses (0.1 s), and $\tau$ is the transit time of the molecules in the ionizing beam volume, i.e., the time that molecules reside in the REMPI beam, and hence the limiting time for the creation of $a^3\Sigma^+$ state molecules that can be ionized. For a 4 mm beam diameter and molecules with kinetic energy ~1 mK, $\tau$ ~ 10 ms. The PA level, $2(0^-)$ $u_{PA} = 5$, could decay to many different vibrational levels of the lowest triplet state $a^3\Sigma^+$. Nevertheless in Fig. 6 we notice that this decay is mainly to the $v'' = 12$ level. In Eq. (4) *FCF* stands for the Franck-Condon factor (FCF) for decay from the PA level to the $v'' = 12$ level of $a^3\Sigma^+$. In this case, *FCF* = 0.2. The FCF is calculated from *ab initio* LiRb PECs [19] using LEVEL 8.0 [26]. Substituting the FCF, transit time $\tau$, and $N_a$ into Eq. (4), we calculate the excited state molecule formation rate to be $R_{PA} = 2\text{x}10^4 - 2\text{x}10^5$ s$^{-1}$. Within the uncertainty in the photoionization cross-section $\sigma$ and detector efficiency $e_d$, this estimate leads to a photoassociation rate higher than other molecular species, for instance, Rb$_2$ [29], consistent with the extremely high $R_{PA}$ observed in trap-loss measurements [16].

## VI. CONCLUSION

We have discussed the formation of ultracold $^7$Li$^{85}$Rb molecules in the lowest triplet state using PA followed by spontaneous emission, and presented preliminary REMPI spectra to probe the high-lying vibrational levels of the lowest triplet state and excited electronic states. Using ionization detection, we have discovered new PA resonances with binding energies up to ~62 cm$^{-1}$. Next, we have recorded REMPI spectra by fixing the PA laser at three different resonances, the $u_{PA} = 5$, 9 and 11 vibrational lines of the $2(0^-)$ electronic state, and identified transitions $3^3\Pi$ ($v_\Pi' = 0 - 10$) $\leftarrow$ $a^3\Sigma^+$ ($v'' = 7 - 13$). Furthermore, we have also labeled lines corresponding to $4^3\Sigma^+$ ($v_\Sigma' = 0 - 5$) $\leftarrow$ $a^3\Sigma^+$ ($v'' = 7 - 13$) – transitions. Overall, we have identified ~94% of the lines experimentally observed in our REMPI spectra and derived the spectroscopic constants for these newly observed electronic states. At this point, we do not have sufficient information to explain the origin of the unassigned lines. They may be connected with one of several other electronic states – besides the two we have studied – that may be accessible by our REMPI laser, as seen in Fig. 3(a). Further investigation in this direction may be warranted. Nevertheless, these initial investigations of the $a^3\Sigma^+$, $3^3\Pi$ and $4^3\Sigma^+$ electronic states of the $^7$Li$^{85}$Rb system are a reasonable starting point for our ongoing work on identifying suitable transition pathways to transfer short-lived ultracold molecules into stable, deeply bound -molecules.

Acknowledgements: We gratefully acknowledge support for early stages of this project from the NSF (CCF-0829918), an equipment grant from the ARO (W911NF-10-1-0243), and university support through the Purdue OVPR AMO incentive grant. The work of JPR was supported by the Department of Energy, Office of Science, under Award Number DE-SC0010545.

# REFERENCES

1. K. Góral, L. Santos and M. Lewenstein, Phys. Rev. Lett. **88**, 170406 (2002).

2. R. V. Krems, Phys. Chem. Chem. Phys. **10**, 4079 (2008).

3. M. G. Moore and A. Vardi, Phys. Rev. Lett. **88**, 160402 (2002).

4. K. –K. Ni, S. Ospelkaus, D. Wang, G. Quéméner, B. Neyenhuis, M. H. G. de Miranda, J. L. Bohn, J. Ye and D. S. Jin, Nature **464**, 1324 (2010).

5. J. J. Hudson, B. E. Sauer, M. R. Tarbutt and E. A. Hinds, Phys. Rev. Lett. **89**, 023003 (2002).

6. E. R. Hudson, H. J. Lewandowski, B. C. Sawyer and J. Ye, Phys. Rev. Lett. **96**, 143004 (2006).

7. C. Chin, V. V. Flambaum and M. G. Kozlov, New J. Phys. **11**, 055048 (2009).

8. D. DeMille, Phys. Rev. Lett. **88**, 067901 (2002).

9. R. Zadoyan, D. Kohen, D. A. Lidar and V. A. Apkarian, Chem. Phys. **266**, 323 (2001).

10. K. Aikawa, D. Akamatsu, M. Hayashi, K. Oasa, J. Kobayashi, P. Naidon, T. Kishimoto, M. Ueda and S. Inouye, Phys. Rev. Lett. **105**, 203001 (2010).

11. S. Ospelkaus, A. Pe'er, K. –K. Ni, J. J. Zirbel, B. Neyenhuis, S. Kotochigova, P. S. Julienne, J. Ye and D. S. Jin, Nature Phys. **4**, 622 (2008).

12. M. Aymar and O. Dulieu, J. Chem. Phys. **122**, 204302 (2005).

13. M. Ivanova, A. Stein, A. Pashov, H. Knöckel and E. Tiemann, J. Chem. Phys. **134**, 024321 (2011).

14. S. Dutta, A. Altaf, D. S. Elliott and Y. P. Chen, Chem. Phys. Lett. **511**, 7 (2011).

15. M. Ivanova, A. Stein, A. Pashov, H. Knöckel and E. Tiemann, J. Chem. Phys. **138**, 094315 (2013).

16. S. Dutta, J. Lorenz, A. Altaf, D. S. Elliott and Y. P. Chen, Phys. Rev. A **89**, 020702(R) (2014).

17. S. Dutta, D. S. Elliott and Y. P. Chen, EPL (Europhys. Lett.) **104**, 63001 (2013).

18. M. Korek, A. R. Allouche, M. Kobeissi, A. Chaalan, M. Dagher, K. Fakherddin and M. Aubert-Frécon, Chem. Phys. **256**, 1 (2000).

19. M. Korek, G. Younes and S. Al-Shawa, J. Mol. Spec: THEOCHEM **899**, 25 (2009).

20. I. Jendoubi, H. Berriche, H. Ben Ouada, and F. X. Gadea, J. Phys. Chem. A **116**, 2945 (2012).

21. S. Dutta, A. Altaf, J. Lorenz, D. S. Elliott and Y. P. Chen, J. Phys. B: At. Mol. Opt. Phys. **47**, 105301 (2014).

22. M. Movre and R. Beuc, Phys. Rev. A **31**, 2957 (1985).

23. A. Ridinger, S. Chaudhuri, T. Salez, D. R. Fernandes, N. Bouloufa, O. Dulieu, C. Salomon and F. Chevy, EPL (Europhysics Lett.) **96**, 33001 (2011).

24. See supplementary material at [URL] for an additional PA resonance and the complete set of REMPI spectra with the line positions and their assignments.

25. R. J. LeRoy and R. B. Bernstein, J. Chem. Phys. **52**, 3869 (1970).

26. R. J. LeRoy, LEVEL 8.0, University of Waterloo Chemical Physics Research Report No. CP-642R (2007). See http://leroy.uwaterloo.ca/programs/

27. T. Bergeman, A. J. Kerman, J. M. Sage, S. Sainis and D. DeMille, Eur. Phys. J. D **31**, 179 (2004).

28. J. Banerjee, D. Rahmlow, R. Carollo, M. Bellos, E. E. Eyler, P. L. Gould and W. C. Stwalley, J. Chem. Phys. **138**, 164302 (2013).

29. M. A. Bellos, D. Rahmlow, R. Carollo, J. Banerjee, O. Dulieu, A. Gerdes, E. E. Eyler, P. L. Gould and W. C. Stwalley, Phys. Chem. Chem. Phys. **13**, 18880 (2011).